\documentclass[referee]{aa} 
\usepackage{graphicx}
\usepackage{txfonts}
\usepackage{xspace}

\newcommand{\prm}[1]{\ensuremath{#1}\xspace}
\newcommand{\tld}{\prm{u}}
\newcommand{\trr}{\prm{k}}
\newcommand{\tip}{\prm{b}}
\newcommand{\tpr}{\prm{p}}
\newcommand{\ttc}{\prm{T_c}}
\newcommand{\twd}{\prm{\Upsilon}}

\begin{document}
   \title{Detection of transit timing variations in excess of one hour in the Kepler multi-planet candidate system KOI 806 with the GTC}

   \author{B. Tingley\inst{1,2}
          \and
          E. Palle\inst{1,2}
          \and
          H. Parviainen\inst{1,2}
          \and
          H.~J. Deeg\inst{1,2}
          \and
          M.~R. Zapatero Osorio\inst{3}
          \and
          A. Cabrera-Lavers\inst{1,2}
          \and
          J.~A. Belmonte\inst{1,2}
          \and 
          P. Monta\~n\'es Rodriguez\inst{1,2}
          \and
          F. Murgas\inst{1,2}
          \and
          I. Ribas\inst{4}
          }

   \institute{Instituto de Astrof\'{i}sica de Canarias,
              C/ V\'{i}a L\'{a}ctea, s/n,
              38205 - La Laguna (Tenerife), Spain\\
              \email{btingley@iac.es}
         \and
              Dpto. de Astrof\'isica,
              Universidad de La Laguna,
              38206 - La Laguna (Tenerife), Spain
         \and
              Centro de Astrobiolog\'ia (CSIC-INTA),
              Crta. de Ajalvir km 4, E-28850,
              Torrej\'on de Ardoz, Madrid, Spain
         \and
              Institut de Ci\`{e}nces de l'Espai (CSIC-IEEC),
              Campus UAB, Fac. de Ciències,
              Torre C5-parell-$2^{\rm \underline{a}}$ planta,
              08193 Bellaterra, Spain
             }

   \date{Received October 15, 2011; accepted XXXXX, 2011}

\abstract
   {}
  {We report the detection of transit timing variations (TTVs) well
    in excess of one hour in
    the Kepler multi-planet candidate system KOI 806. This system exhibits transits consistent
    with three separate planets -- a Super-Earth, a Jupiter, and a
    Saturn -- lying very nearly in a 1:2:5 resonance, respectively.}
   {We used the Kepler public data archive and
  observations with the Gran Telescopio de Canarias to compile the necessary
  photometry.}
{For the largest candidate planet (KOI 806.02) in this system, we
  detected a large transit timing variation of -103.5$\pm$6.9 minutes
  against previously published ephemeris. We did not obtain a
    strong detection of a transit color signature consistent with a
    planet-sized object; however, we did not detect a color difference
    in transit depth, either.}
  {The large TTV is consistent with theoretical predictions that
    exoplanets in resonance can produce large transit timing
    variations, particularly if the orbits are eccentric. The
      presence of large TTVs among the bodies in this systems
      indicates that KOI806 is very likely to be a planetary
      system. This is supported by the lack of a strong color
      dependence in the transit depth, which would suggest a blended
      eclipsing binary.}

\titlerunning{Detection of TTVs in KOI 806}

   \maketitle

\section{Introduction}

Transit timing variations (TTVs) are proving to be a very valuable
tool in exoplanet research. In systems with transits of
different periodicities, they may serve to constrain the mass of the
transiting bodies and thereby verify them as planets, without further
observations by e.g. radial velocities. Furthermore, TTVs may lead to
the discovery of further non-transiting planets.
Indeed, TTVs have been instrumental in characterizing multiple systems
found by Kepler, including Kepler-9 (Holman et al. \cite{hol2011}),
Kepler-11 (Lissauer et al. \cite{lis2010}), Kepler-16 (Doyle et
al. \cite{doy2011}), Kepler-18 (Cochran et al. \cite{coc2011}), and
Kepler-19 (Ballard et al. \cite{bal2011}), where an additional,
non-transiting planet has been identified solely through TTVs.

\begin{figure}
\centering
\includegraphics[width=\columnwidth]{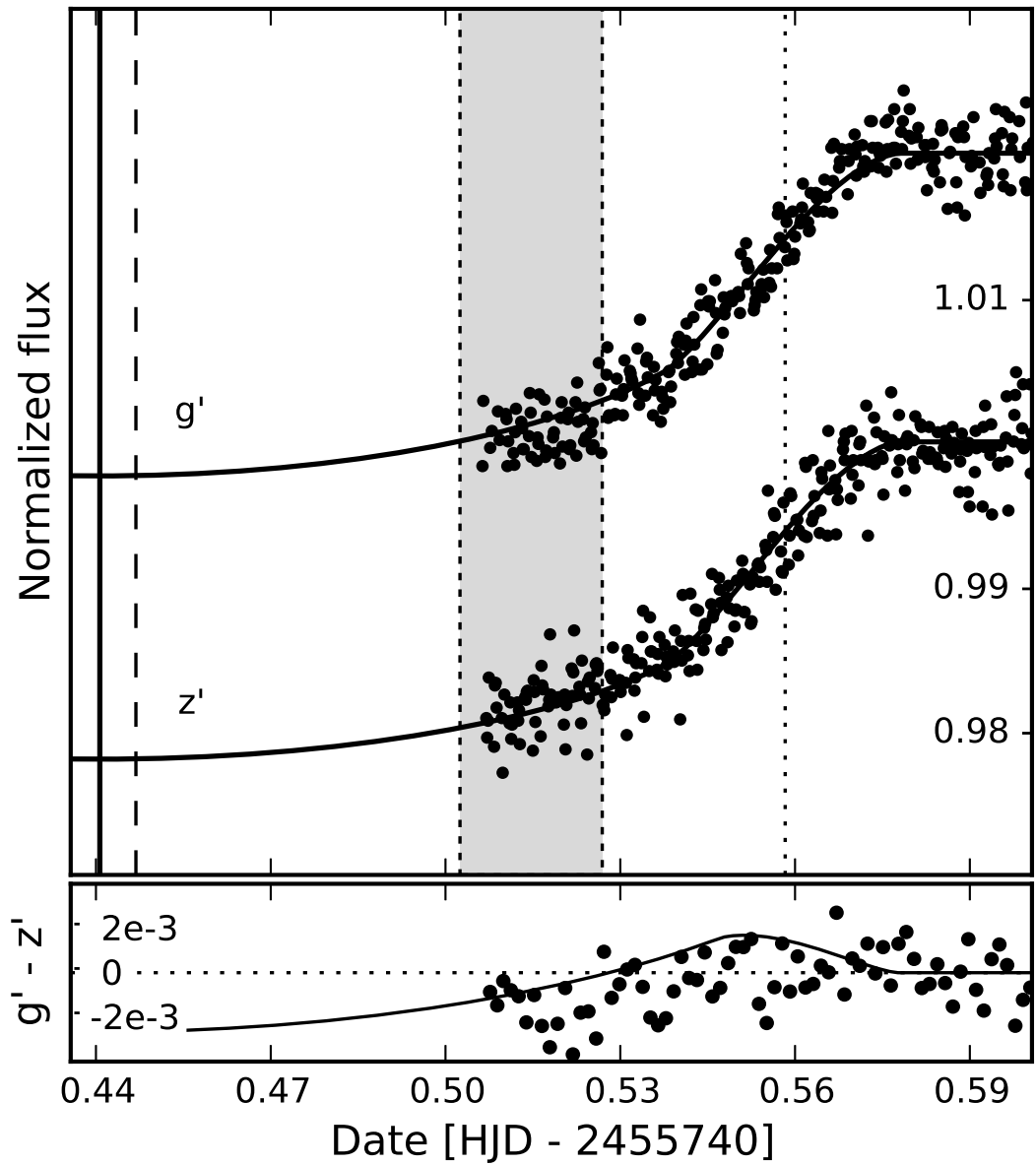}
\caption{GTC photometry of a transit of candidate KOI 806.02 in $g'$
  (top) and $z'$ (bottom). The solid line is the fitted transit center
  and the dashed line its 3$\sigma$ confidence limit, the shaded
  region defines the transit center $\pm 3\sigma$ predicted by B11,
  and the dotted line is mid-way through the egress. The time of
  transit center predicted by the ephemeris of B11 is $2455740.5146
  \pm 4.2\times10^{-3}$, about 1.7$h$ earlier. The lower figure shows the
  observed $g'-z'$ during the transit egress, along with a
  modeled color signature using limb darkening models from Claret \&
  Bloemen (\cite{cla2011}). The points shown are the mean of each $4\times5s$
  exposure sequence.}
\label{gtc806}
\end{figure}


\begin{figure}
\centering
\includegraphics[width=\columnwidth]{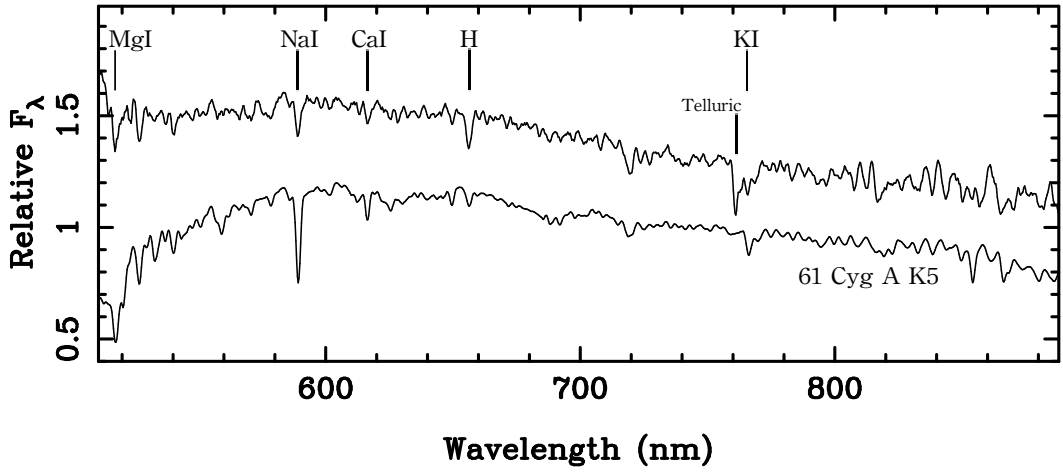}
\caption{ALFOSC/NOT spectrum for KOI 806 (top) along with a K5 standard
  star observed with the same instrumental configuration on the same
  night. The two spectra are smoothed by 5 pixels and offset by 0.3
  for clarity, with the pertinent atomic lines indicated.}
\label{spectrum}
\end{figure}

\begin{figure}
\centering
\includegraphics[width=\columnwidth]{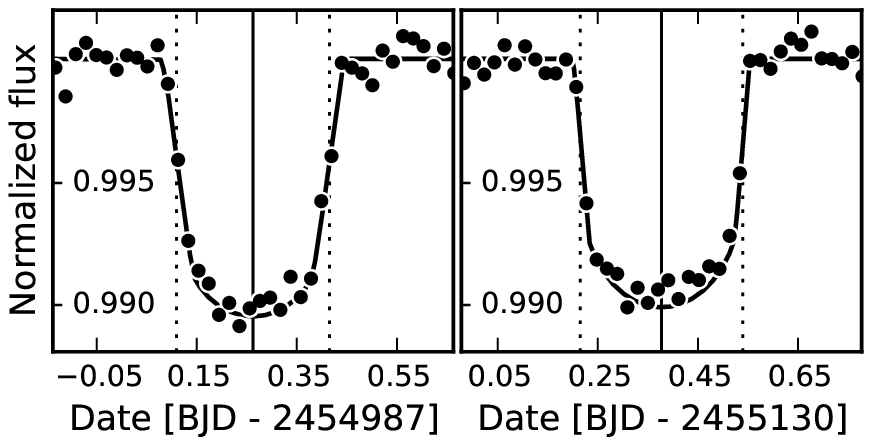}
\includegraphics[width=\columnwidth]{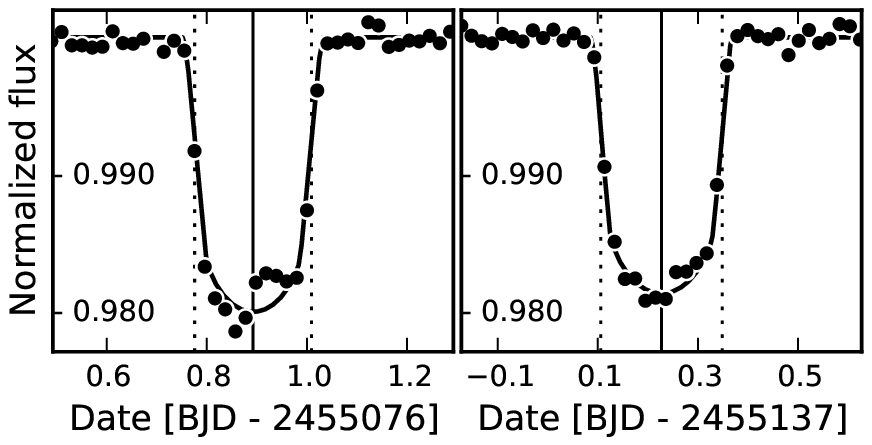}
\caption{Public Kepler photometry for the two released transits of
  KOI 806.01 (top), KOI 806.02 (bottom) along
  with the best fit transits.}
\label{keplerobs}
\end{figure}


\begin{figure}
\centering
\includegraphics[width=\columnwidth]{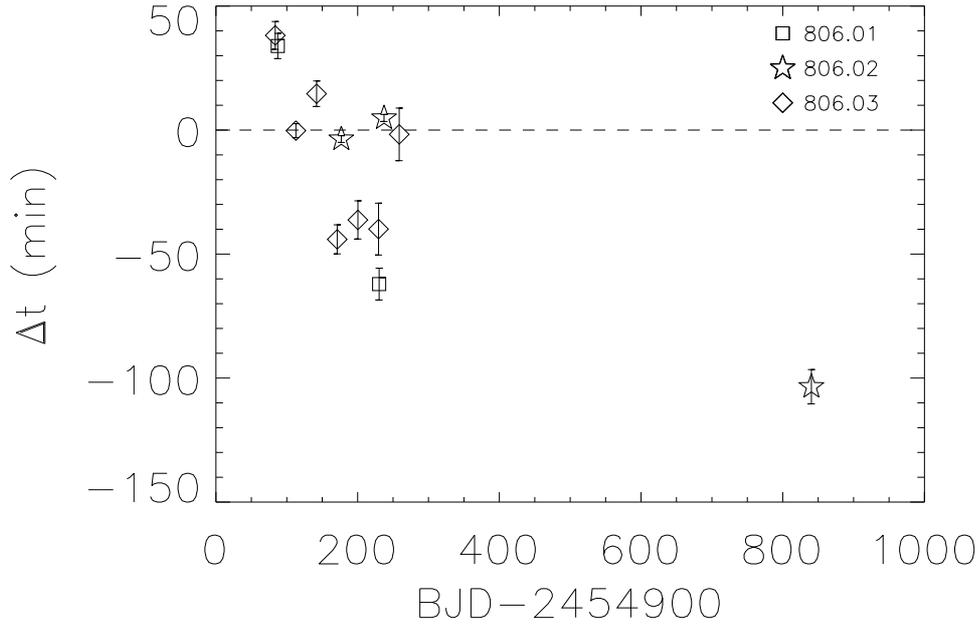}
\caption{Transit timing variations for the KOI 806 candidates vs. the
  periods and $T_0$s given in B11. All of the candidates in the system
  exhibit high significance ($\sigma > 6$) TTVs, with KOI 806.01 and
  KOI 806.03 doing so in just the Kepler Q0-Q3 data. KOI 806.01 has a
  min/max difference of over 1.5 hours for just the first two
  transits, KOI 806.02 1.8 hours, and KOI 806.03 1.4 hours. These values
  are all tabulated in Table~\ref{ttvtable}.}
\label{ttvplot}
\end{figure}

Since its public release in Feb. 2011 of the list of Kepler planet
candidates (Borucki et al. \cite{bor2011}, hereafter B11), the system
KOI 806 has been an obvious candidate for large TTVs. The innermost
two planet candidates, designated 806.03 and 806.02, (hereafter '{\em
  03}' and '{\em 02}') are given with periods of 29.1654$\pm$0.0012
days and 60.32875$\pm$0.00037, respectively, and a further candidate,
806.01 (hereafter '{\em 01}'), is listed with $P$=143.1814$\pm$0.0027
days.  The periods of {\em 02} and {\em 03} lie very close to the 2:1
resonance, which can strongly enhance TTVs relative to a
  non-resonant state (e.g. Agol et al. \cite{ago2005}). However, the
analysis of transit timing observations from Kepler observing segments
Q0 - Q2 (corresponding to data up to JD 2455091.48) by Ford et
al. (\cite{for2011}) stops short of listing this system among clear
candidates for detected TTVs. The apparent reason is that in data from
Q0-Q2, only the shortest-periodic {\em 03} exhibited possible TTV's,
whereas {\em 01} and {\em 02} had only one transit in these
segments. Even so, the TTVs predicted by Ford et al. for {\em all}
candidates in KOI 806 are among the highest of the candidates in B11;
Ford et al. predict that the min-to-max TTVs over 3.5 years could
range from $\sim 15$ minutes for {\em 01} to almost 5 hours for {\em
  03}, the smallest candidate with the shortest orbit.

\section{Observations}

We used the 10.4m Gran Telescopio Canarias (GTC) to observe a transit
of candidate {\em 02} with semi-simultaneous time series photometry in
 $g'$ and $z'$, 2 widely separated colors chromatically. These
observations were undertaken in an attempt to verify that the
transiting body is indeed of planetary size, as it should then show
characteristic color variations due to the interplay between the
relative size of the transiting body and differential stellar limb
darkening (Tingley \cite{tin2004}). While such detection of the color
``signature'' of an exoplanet can be difficult, {\em 02} has a deep
transit (almost 2\%) and a long transit duration (6.6 hrs), which made
it an excellent candidate for the attempt. We obtained approximately
1800 images with GTC/OSIRIS in five hours on the night of the 27/28th
of June, 2011, alternating four 5-second exposures in $g'$ ($468.8 \,
nm$) with four 5-second exposures in $z'$ ($893.1\, nm$). Subsequent
analysis using optimized aperture photometry with the 'Vaphot' package
(Deeg \& Doyle \cite{dee2001}) for IRAF showed that the transit 
  unexpectedly ended $\approx$ 100 {\em min} earlier than anticipated
  by the period and duration given in B11, a detection of a TTV with a
  significance of $\sim15\sigma$. Given the presence of 3 candidate
  transits in this system and the high significance of this TTV
  detection, it is extremely likely, just based on this, that KOI 806
  harbors a planetary system of at least three planets.

In the light of this very strong TTV detection, the original science
goal based on multi-color photometry becomes relegated to supporting
evidence. Despite the TTV, we
captured the crucial part of the egress where the planet crosses the
limb of the star, which results in a weak detection of a $g'-z'$ color
signature (on the order of one millimag), consistent with the modeled
one, though offset (Fig.~\ref{gtc806}). This offset is plausible,
given the clear evidence of transparency variations just after the
transit, revealed by increased scatter in the photometry. This could
in turn affect the measured out-of-transit flux.

However, even without a true detection of the transit color signature,
we can constrain the the nature of any false positive. The photometry
does clearly excludes the possibility of any strong dependency of
transit depth on color, which one would expect from a blend with any
color difference between the components, as proposed by Tingley
(\cite{tin2004}) and utilized by O'Donovan et al.  (\cite{odo2006}),
O'Donovan et al. (\cite{odo2007}), Cochran et al. (\cite{coc2011}),
and Ballard et al. (\cite{bal2011}). We derived an equation that is
similar in principle to BLENDER (Torres et al \cite{tor2004,tor2011}),
but much simplified, to limit the possible color difference between
the target star and any contaminating eclipsing binaries (CEBs). It
relates the blended eclipse depth in $z'$ ($d_{CEB,g}$) to that in $g'$
using only the depth of the unblended eclipse depth in $g'$ and $z'$ (
$d_{EB,g}$ and $d_{EB,z}$, respectively) and the $g'-z'$ color difference
between the target star and the blending eclipsing binary ($\Delta(g'-z')$):

\begin{equation}
d_{CEB,z} = \frac{1 + d_{EB,z} f_{EB,z}}{1+f_{EB,z}},
\end{equation}
where
\begin{equation}
f_{EB,z} =
\frac{1-d_{CEB,g}}{d_{CEB,g}-d_{EB,g}}f_{target,g}10^{0.4\Delta(g-z)}
\end{equation}
where we set $f_{target,g}=1$, which yields a differential blended
eclipse depth in $z'$. If we set $d_{EB,z} = d_{EB,g}$,
$\Delta(g'-z')$ must be less than about 0.02. Relaxing this constraint
to allow $d_{EB,z}$ to vary as much as 10\% from $d_{EB,g}$, a
reasonable value, constrains $\Delta(g'-z')$ to about 0.25 for lightly
blended eclipses ($f_{target} \sim f_{EB}$) to 0.05 for highly blended
eclipses ($f_{target} \gg f_{EB}$). While this does not absolutely rule
out the possibility that the transit of {\em 02} may be due to a blend,
it does suggest that it is unlikely.


While the GTC showed a clear advance of the transit egress, we
  would not have been able to detect any changes in transit duration
  due to the partial coverage of the transit. To verify the stellar
parameters given in the Kepler Input Catalogue for the host star, KIC
3832474 ($T_{\rm eff} = 5206 K$, $\log g = 4.53$), we took
spectra of KOI 806 and several other similar stars with the 2.5 meter
Nordic Optical Telescope (NOT), using grism \#5 at a resolving power
R=830 over the wavelength interval 5100-9000\AA. The spectra were
calibrated in wavelength against Ne lines from arc images taken with
the same instrumental configuration as the target and flux normalized
to unity at the wavelength interval 7400-7500\AA. Uncertainty in
instrumental response correction is less than 10\%. We corrected the
spectra for telluric absorption, though some residuals at the
strongest telluric oxygen band at around 7600\AA \, remain due to
different airmasses of target and telluric standard star (see
Fig.~\ref{spectrum}). We obtained the spectral classification of the
host star via atomic line identification (MgI, NaI, H$\alpha$, and
CaII), which indicated a spectral type of G9-K0$\pm$2 and a slightly
higher temperature ($T_{eff} \sim5250$K) than given in B11. We did not
have the necessary spectral precision to estimate $\log g$.

This leaves the question of the radius of the star open, given by B11
as $R_\star=0.88 R_\odot$. A Jupiter-sized planet transiting a star of
this size would cause a transit with a depth of about 1.4\%, whereas
the observed transits of {\em 02} have a depth of 2.0\%. Therefore,
for the transits of {\em 02} to be due to a planet, {\em 02} needs to
be $\sim$ 20\% larger than Jupiter or the star has to be smaller than
expected by a similar amount -- or, most likely, some combination of
the two.

\begin{table*}
\caption{Best-fit transit times for the planet candidates KOI 806.01,
KOI 806.02 and KOI 806.03, and their offset (O-C) against the ephemeris given in B11. $N$ is the transit number and the dates are in BJD-2454900.}
\label{ttvtable}
\centering
\begin{tabular}{c c c | c c c}
\hline\hline
$N$ & $T_{c,fitted}$ & O-C (min) & $N$ & $T_{c,fitted}$ & O-C (min)\\
\hline
  & KOI 806.01             &                &
  & KOI 806.03            &              \\
0 & $87.26254 \pm 5.4\times10^{-4}$  & $33.9 \pm 5.1$ &
0 & $83.7185 \pm 3.8\times10^{-3}$  & $38.2 \pm 5.6$\\
1 & $230.37726 \pm 5.7\times10^{-4}$ & $-62.1\pm 6.4$ &
1 & $112.8572 \pm 1.3\times10^{-3}$  & $-0.2 \pm 2.8$\\
 &                          &                &
2 & $142.0330 \pm 2.5\times10^{-3}$  & $14.7 \pm 5.1$\\
  & KOI 806.02            &                 &
3 & $171.1576 \pm 1.7\times10^{-3}$  & $-44.1 \pm 5.9$\\
0 & $176.89249 \pm 3.5\times10^{-4}$  & $-3.6 \pm 1.4$ &
4 & $200.3284 \pm 2.2\times10^{-3}$   & $ -36.2 \pm 7.7$\\
1 & $237.22714 \pm 2.3\times10^{-4}$  & $4.9 \pm 1.4$ &
5 & $229.4913 \pm 4.0\times10^{-3}$   & $-39.9 \pm 10.5$\\
11$^*$& $840.4394 \pm 2.4\times10^{-3}$   & $-103.5 \pm 6.9$ &
6 & $258.6832 \pm 1.2\times10^{-3}$   & $-1.7 \pm 10.6$\\
\hline
$^*$From egress observed by GTC\\ 
\end{tabular}
\end{table*}

\section{Analysis of transits}

With the recent release of Kepler Q3 data, covering up to 16 Dec 2009
(JD 2455182.00), we were able to extend our analysis, as second
transits of both {\em 01} and {\em 02} are now available. We performed
fits (see discussion below) to the transits in Kepler data of each the
three candidates (see Fig.~\ref{keplerobs} for {\em 01} and {\em 02},
{\em 03} similar but not shown).

Our parameter estimation process is based on a Bayesian
methodology. We first search for the rough global maximum likelihood
solution using the differential evolution (DE) global optimization
method (Price, Storn \& Lampinen \cite{pri2005}; Storn \& Price
\cite{sto1997}), and then derive the parameter posterior distributions
using Markov Chain Monte Carlo (MCMC) (Ford \cite{for2005}).  We
parameterize the transits by the transit center \ttc, squared radius
ratio \trr, period \tpr, impact parameter \tip, and reciprocal of the
half duration of the transit \twd, similar to Kipping
(\cite{kip2010b}) and Bakos et al. (\cite{bak2010}). Further, we use the
linear limb darkening law with a free limb darkening coefficient \tld.

We use unconstraining (uninformative) priors for the parameters
estimated for the two largest candidates, {\em 01} and {\em 02}. For
the smallest candidate, {\em 03}, we use normal distribution priors to
constrain the values allowed for \tip, \twd and \tld.  The prior
centers for \tip and \twd are adopted from the released Kepler values
and widths are chosen conservatively, as we prefer to overestimate the
errors slightly rather than underestimate them. The limb darkening
prior is derived from the limb darkening posteriors of the fits to
{\em 01} and {\em 02}.

For the fits to the long-cadence Kepler data, we subsample our model
by 10 (i.e., each final model light curve point is an average of 10
model points uniformly spread over the time of a single Kepler
exposure). This is necessary since long exposures modify the transit
shape (Kipping \cite{kip2010c}). The Kepler observations and the
corresponding fits can be seen in Fig.~\ref{keplerobs} for {\em 01}
and {\em 02}.

The fit to our GTC observations of candidate {\em 02} is done
simultaneously in both colors. The only major difference to the single
color Kepler data is that we have separate limb darkening coefficients
for each color. Since we have only a partial transit, we cannot infer
the transit duration. Instead, we use an informative prior of $5.7 \pm
0.1 h$ from the transit duration posteriors of the Kepler transit
fits, and similarly, a prior for the impact parameter of
\tip$=0.550\pm0.075$. The rest of the parameters have uninformative
priors for conservative error estimates. The resulting residuals have
a $\sigma_{\rm rms}$ of 1.9 millimags in $g'$ and 2.1 millimags in $z'$
per $4\times5$ exposure sequence.

With the time
of the transit center being one of the free fit parameters, we were
able to measure the periods from the first two transits of {\em 01}
and {\em 02} and the first seven transits of {\em 03}, respectively,
for comparison to the period given by B11. In combination with our own
observations with the GTC, we can confirm that large and statistically
very significant TTVs are apparent in this system (Fig.~\ref{ttvplot}
and Table~1).

\section{Conclusions}

We have detected strong TTVs in the KOI 806 system, affecting the
timing of the transit far more than what we had expected based on the
uncertainties indicated by the ephemeris of B11. The magnitude of the
TTVs seen in this system are in fact not surprising, as Holman \&
Murray (\cite{hol2005}) have argued that TTVs may scale in a roughly
linear fashion with period, though TTVs are highly degenerate and
strongly dependent on initial conditions (Nesvorn\'y \& Morbidelli
\cite{nes2008}, Veras, Ford \& Payne \cite{ver2011}). If this
  linear relationship holds, then the observed TTVs in this system are
  similar in scale to those observed in, for example, the Kepler-11
  system (Lissauer et al. \cite{lis2010}) - with TTVs of tens of
  minutes at periods of 10 - 47 $d$. Our observations of {\em 02}
  indicated a TTV of $-103.3 \pm 6.9\, min$ relative to the transit
  ephemeris given by the Kepler candidate list (B11). These values are
  based on a linear fit to all the transit that occurred during Kepler
  quarters Q0-Q5. Over this period of time, {\em 02} transited 5
  times, yet the B11 values imply a transit ephemeris error of only
  $\sim 4\, min$ for the 12th transit, the one we observed with GTC
  (see Fig.\ref{gtc806}). We performed an independent analysis of the
  publicly available transits (Q0-Q2), finding significant differences
  to the periods and $T_0$s given in B11. For example, we derive a
  period of $60.33465\pm0.00042\, d$ for {\em 02}, which results in a
  TTV for the 12th transit of $\sim 3.2 \, h$.  However, as much of
  the Q0-Q5 data are still proprietary, we cannot perform a full,
  independent revision of the method and results that led to the
  ephemeris given by B11. A TTV well in excess of one hour is certain,
  though, regardless of which period is used, either ours from the
  first two transits or that from B11. The second result of our GTC
  observations is the absence of any color dependency of the transit
  depths, the presence of which would reject the candidate, along with
  a color signature during egress that suggests that an object with a
  radii consistent with a planet is causing the transit, albeit with a
  low statistical significance.

While our data do not allow for a detailed analysis of the system, we
consider that the existence of a very strong TTV and the absence of a
color dependency of the transit depth strongly supports that the
transits in KOI 806 are caused by planetary companions.  Two possible
scenarios for massive companions exist: either the planet is orbiting
one star or is circumbinary. Transiting circumbinary planets have a
complex pattern of transits due to the relatively large motion of the
stellar components as they orbit around the center of mass, which is
clearly not seen is this system (Deeg et al. \cite{dee1998}, Doyle et
al. \cite{doy2011}). TTVs in systems where the planet orbits one star,
called S-type orbits, are caused by the light time effect (e.g. Kopal
\cite{kop1959}, as distance between the transited star and the
observer changes as the stars orbit around the center of mass. While
this can cause large TTVs over time, a 1.7 $h$ TTV requires an
acceleration amounting to a change of about 13 AUs. Such a change
could only be explained by some exotic, invisible and very massive
object (e.g. a black hole), as a small, faint star could not move
  the primary star so far in the less than 3 years over which the TTVs
  have been measured.  and is therefore very unlikely. Moreover, it
could still not explain the TTVs measured in the other planets in the
system, assuming the values in B11 are correct.

Given the long periods of the putative planets in KOI 806, it is
entirely possible that this system will not be fully characterized by
the end of the Kepler mission -- particularly if evidence of other
  unknown, perhaps longer period planets arises. However, the deep
transits and large TTVs make it possible to monitor this system from
ground over the long term -- an endeavor likely to be scientifically
worthwhile, depending on the details that emerge from the analysis of
the data currently being acquired by the Kepler mission.

A final aspect of unknown TTVs in this and other planetary systems
pertains to photometric follow-up. It appears that the transit
ephemeris given in B11 do not (or cannot) necessarily account for TTVs
in all cases, resulting in ephemeris errors that significantly
underestimate the true timing error. Therefore, observers intending to
follow up Kepler candidates, particularly in known multi-candidate
systems where large TTVs may be present, need to include this fact
when planning their observations.

\begin{acknowledgements}
This article is based on observations made with the GTC operated on
the island of La Palma by the IAC in the Spanish Observatorio of El
Roque. The Kepler data presented in this paper were obtained from the
Multimission Archive at the Space Telescope Science Institute
(MAST). STScI is operated by the Association of Universities for
Research in Astronomy, Inc., under NASA contract NAS5-26555. Support
for MAST for non-HST data is provided by the NASA Office of Space
Science via grant NNX09AF08G and by other grants and contracts. BT and
HD acknowledge funding by grant AYA2010-20982-C02-02 and MRZO by grant
AYA2010-21308-C03-02, both of the Spanish Ministry of Science and
Innovation (MICINN). HP is supported by RoPACS, a Marie Curie Initial
Training Network funded by the European Commission's Seventh Framework
Programme.

\end{acknowledgements}

\end{document}